\documentclass[preprint,aps,showpacs]{revtex4}
\usepackage{graphicx}
\usepackage{epstopdf}
\usepackage{dcolumn}
\usepackage{bm}
\usepackage{color}

\begin{document}


\title{Graphene on Rh(111): STM and AFM studies}

\author{E. N. Voloshina,$^1$ Yu. S. Dedkov,$^{2,}$\footnote{Corresponding author. E-mail: Yuriy.Dedkov@specs.com} S. Torbr\"ugge,$^{2}$ A. Thissen,$^2$ and M. Fonin$^3$}
\affiliation{$^1$Physikalische und Theoretische Chemie, Freie Universit\"at Berlin, 14195 Berlin, Germany}
\affiliation{$^2$SPECS Surface Nano Analysis GmbH, Voltastra\ss e 5, 13355 Berlin, Germany}
\affiliation{$^3$Fachbereich Physik, Universit\"at Konstanz, 78457 Konstanz, Germany}

\date{\today}

\begin{abstract}

The electronic and crystallographic structure of the graphene/Rh(111) moir\'e lattice is studied via combination of density-functional theory calculations and scanning tunneling and atomic force microscopy (STM and AFM). Whereas the principal contrast between \textit{hills} and \textit{valleys} observed in STM does not depend on the sign of applied bias voltage, the contrast in atomically resolved AFM images strongly depends on the frequency shift of the oscillating AFM tip. The obtained results demonstrate the perspectives of application atomic force microscopy/spectroscopy for the probing of the chemical contrast at the surface.
\end{abstract}


\maketitle

Graphene, a single layer of carbon atoms ordered in a ``chicken-wire'' lattice~\cite{Geim:2009,Geim:2011,Novoselov:2011}, is proposed to be used in many technological applications. Among them are gas sensors~\cite{Schedin:2007}, THz-transistors~\cite{Lin:2010}, integrated circuits~\cite{Lin:2011}, touch screens~\cite{Bae:2010}, and many others~\cite{Geim:2009,Novoselov:2011}. One of the promising systems on the basis of graphene is its interface with  metallic substrates~\cite{Wintterlin:2009,Batzill:2012,Dedkov:2012book}. Here graphene can be used as a protection layer for the underlying substrate~\cite{Dedkov:2008d,Dedkov:2008e,Sutter:2010bx,Chen:2011a}, as a spin-filtering material separating two layers of a ferromagnetic material~\cite{Karpan:2007,Karpan:2008,Dedkov:2010a}, or, in case of its growth on a lattice mismatched surfaces [for example, Ir(111), Rh(111), or Ru(0001)], as a template for the preparation of ordered arrays of clusters~\cite{NDiaye:2009a,Sicot:2010}.

In the row of the graphene/metal lattice mismatched systems, the graphene/Rh(111) interface can be considered as an intermediate case between two systematically studied graphene-metal systems: graphene/Ir(111) and graphene/Ru(0001).  These three interfaces represent the situation of a relatively large lattice mismatch between graphene and a metallic substrate. As was demonstrated, graphene on Ir and Ru can be considered as two extreme cases of weakly and strongly interacting interfaces, respectively. For graphene on Rh(111) [see Fig.\,1(a)] several regions of different arrangements of carbon atoms above a Rh(111) substrate can be found. When using the common notations, the following high-symmetry places can be identified: $ATOP$ [A; carbon atoms are placed above Rh(S-1) and Rh(S-2) atoms], $HCP$ [H; carbon atoms are placed above Rh(S) and Rh(S-2) atoms], $FCC$ [F; carbon atoms are placed above Rh(S) and Rh(S-1) atoms], and $BRIDGE$ [B; Rh(S) atoms bridge the carbon atoms]. These places are marked in Fig.\,1(a) by circle, down-triangle, square, and stars, respectively. Among them, the \textit{BRIDGE} positions are expected to be the most energetically favorable for the nucleation of deposited atoms on top of a graphene layer. The nowadays available force spectroscopy and microscopy can shed light on this problem and can be used as a tool, which helps to optimize the preparation of ordered arrays of clusters on a graphene template~\cite{Sugimoto:2007}.

In this manuscript we present the combined study of the graphene/Rh(111) system via application of the state-of-the-art density-functional theory (DFT) calculations and scanning tunneling and atomic force microscopy (STM and AFM). The calculated imaging contrast for STM between all high-symmetry positions for graphene/Rh(111) is in very good agreement with experimental results and this contrast does not depend on the sign of the bias voltage applied between a tip and the sample. As opposed to the latter observation, the imaging contrast in atomically-resolved AFM measurements depends on the frequency shift of the oscillating tip, that can be understood on the basis of measured force-spectroscopy curves. The presented results are compared with the available theoretical data.

The presented in Fig.\,1(a) crystallographic model of graphene/Rh(111) was used in DFT calculations, which were carried out using the projector augmented wave method~\cite{Blochl:1994}, a plane wave basis set with a maximum kinetic energy of $400$\,eV and the PBE-GGA exchange-correlation potential~\cite{Perdew:1996}, as implemented in the VASP program~\cite{Kresse:1994}. The long-range van der Waals interactions were accounted for by means of a semiempirical DFT-D2 approach proposed by Grimme~\cite{ Grimme:2006}. The studied system is modeled using supercell, which has an ($11\times11$) lateral periodicity and contains one layer of ($12\times12$) graphene on four-layer slab of metal atoms. Metallic slab replicas are separated by ca.\,$18$\,\AA\ in the surface normal direction. To avoid interactions between periodic images of the slab, a dipole correction is applied~\cite{Neugebauer:1992}. The surface Brillouin zone is sampled with a single $k$-point at the $\Gamma$ point for structure optimization and set to $3\times3$ in the total energy calculations. The STM images are calculated using the Tersoff-Hamann formalism~\cite{Tersoff:1985}, in its most basic formulation, approximating the STM tip by an infinitely small point source~\cite{VanPoucke:2008,Voloshina:2011NJP}.

The graphene/Rh(111) system was prepared in ultra-high vacuum station for STM/AFM studies according to the recepie described in details in Refs.~\cite{Sicot:2010,Sicot:2012}. The quality and homogeneity were verified by means of low-energy electron diffraction (LEED) and STM. The STM/AFM images were collected with Aarhus SPM 150 equipped with KolibriSensor$\textsuperscript{\texttrademark}$ from SPECS~\cite{SPECS,Torbrugge:2010} with Nanonis Control system. In all measurements the sharp W-tip was used which was cleaned \textit{in situ} via Ar$^+$-sputtering. In presented STM images the tunneling bias voltage, $U_T$, is referenced to the sample and the tunneling current, $I_T$, is collected by the tip, which is virtually grounded. During the AFM measurements the sensor was oscillating with the resonance frequency of $f_0=1001541$\,Hz and the quality factor of $Q=32323$, and the frequency shift was used as an input signal in a feedback loop for the topography measurements. The oscillation amplitude was set to $A=300$\,pm. The base vacuum was better than $8\times10^{-11}$\,mbar during all experiments. All measurements were performed at room temperature.

The DFT-D2 optimized structure is presented in Fig.\,1(a) and the variation of the hight of the carbon atoms is shown in Fig.\,S1 of the Supplementary material~\cite{SUPL}. Carbon atoms in the $ATOP$ configuration define a high-lying region sitting at $d_0=3.15$\,\AA\ above Rh(111), and those in other configurations form a lower region. The buckling in the graphene overlayer is $1.07$\,\AA. Carbon atoms in the $BRIDGE$ configuration form the lowest topographic area ($d_0=2.08$\,\AA). The $HCP$ and $FCC$ regions are approximately $0.4$\,\AA\ and $0.8$\,\AA\ higher than the minima.

We have estimated influence of dispersion forces on the obtained results: while qualitatively the observed picture remains the same, non inclusion of the van der Waals interactions (i.\,e. standard DFT-PBE treatment) yields larger corrugation ($\approx1.8$\,\AA) with a very similar low region ($d_0=2.10$\,\AA), but a high region at $d_0=3.90$\,\AA. This is due to the alternating ``weak'' and ``strong'' interactions of graphene with the Rh(111) surface. In the case of the ``strong'' interaction between graphene and metal, standard GGA-treatment gives reasonable result, whereas for the areas of the ``weakly'' bonded graphene dispersion forces, neglected by the standard procedure, are important.

Fig.\,1(b) shows the DFT-D2-calculated STM image (distance between graphene and the tunneling tip is $z=2$\,\AA) of the graphene layer on Rh(111). All structural replicas are clearly visible in calculated STM image and they are marked by the corresponding symbols in Fig.\,1(b). The brightest region in this image is the one surrounding the $ATOP$ (A; circle) high symmetry arrangement for graphene on Rh(111). Here the graphene layer is most weakly bounded to the Rh substrate. The next two places in the bonding row are $FCC$ (F; rectangle) and $HCP$ (H; triangle). The darkest place in the calculated STM image is the $BRIDGE$ (B; star) position of the carbon atoms above Rh(111), where the strongest interaction of graphene and Rh(111) is expected and was calculated.
 
Fig.\,1(c) and (d) show the experemental STM images of the graphene/Rh(111) system collected over large area and several unit cells of the graphene layer on Rh(111), respectively. The LEED picture of this system is shown as an inset of (c). The results, presented in (d) are the part of the results shown in Fig.\,S2 of the Supplementary material~\cite{SUPL}. The presented data demonstrate the high quality of the studied system over large regions as well as locally and they are in very good agreement with the previously published results~\cite{Sicot:2010,Sicot:2012,Wang:2010ky}. The brightest areas are the $ATOP$ (A) positions and the darkest are the $BRIDGE$ (B) ones. The clear and unique assignment of all arrangements of carbon atoms in the graphene layer and the Rh substrate can be made via comparison of experimental and theoretical data due to the extremely good agreement between them (Fig.\,1 and Fig.\,S2 of the Supplementary material~\cite{SUPL}). The corrugation of the graphene layer on Rh(111) was measured between $0.5$\,\AA\ and $1.5$\,\AA\, depending on the imaging conditions. However, the principal contrast in STM images is not changed upon variation of the tunneling conditions (see Fig.\,S3 in the Supplementary material for the corresponding images~\cite{SUPL}).

The more intriguing results were obtained by AFM imaging of the graphene/Rh(111) system and they are compiled in Fig.\,2 and demonstrate the clear atomic contrast in atomic force microscopy images of graphene on Rh(111). In the beginning, the force spectroscopy of the studied surface was performed via monitoring the frequency shift of the oscillating tip as a function of the distance between the tip and the surface, $\Delta f(z)$. At the same time the tunneling current, $I(z)$, was collected. The representative curves for the two most contrast areas of the graphene/Rh(111) topography, namely $ATOP$ and $BRIDGE$, are shown in the upper panel of Fig.\,2 as open circles and squares, respectively. The corresponding force curves, $F(z)$, are shown in Fig.\,S4 of the Supplementary material~\cite{SUPL}. The tunneling current (shown as a semi-logarithmic plot in the inset of the upper panel of Fig.\,2) demonstrates the exponential dependence on the distance between sample and the tip characteristic for two metallic electrodes separated by vacuum. The horizontal dashed lines indicate the frequency shifts for the KolibriSensor$\textsuperscript{\texttrademark}$ at which the AFM images were subsequently collected. These images with the pointing of the high symmetry stacking positions of graphene/Rh(111) are presented as a low row of Fig.\,2 and marked by the corresponding letters.

The presented force curves can be clearly distinguished from each other that indicates the sensitivity of AFM to the different adsorption sites on the graphene layer on Rh(111). This difference in adsorption energies for different places in the graphene/Rh(111) lattice was pointed earlier in Ref.~\cite{Wang:2011hh}: it was found that the $ATOP$ position is less energetically favorable for adsorption and there is a tendency for atoms to nucleate around the $BRIDGE$ positions. These theoretical observations are supported by our experimental results: the expected equilibrium distance between tip and the sample for the $BRIDGE$ position is smaller compared to the one for the $ATOP$ position and the attractive part of the force [proportional to the frequency shift, $\Delta f(z)$] is large for the $BRIDGE$ position. Thus, AFM imaging of the surface can be used  as a tool for precise detecting of the chemical and the electronic contrasts.

The interesting results are obtained when collecting AFM images at the frequency shifts marked in Fig.\,2: increasing of the frequency shift during imaging leads to approaching the scanning tip closer to the studied surface that yields the increasing imaging contrast. This effect is clearly visible in Figs.\,2(a-c). In (a) one can see the contrast only between ``hills'' and ``valleys'' of graphene/Rh(111), whereas in AFM images shown in (b) and (c) the evident difference between $ATOP$, $HCP$, $FCC$, $BRIDGE$ positions (they are marked by the corresponding letters) as well as atomically-resolved contrast in the grapehene layer are visible. In all these AFM images [Fig.\,2(a-c)] the topography of the graphene/Rh(111) lattice as well as all hight variations are well reproduced. For the discussion of the observed contrast between $HCP$, $FCC$, $BRIDGE$ positions in the obtained AFM images [Fig.\,2(d)] we can refer to Ref.~\cite{Wang:2011hh} (see also Figs.\,S6 and S7 of the Supplementary material~\cite{SUPL} for the calculated site-projected density of states for carbon atoms and the difference of electron density at the graphene/Rh(111) interface, respectively), where the bonding of atoms of different nature on different adsorption sites of the graphene/Rh(111) is discussed. However, the more careful modeling and the further extended discussion is necessary here, that could be a topic of the future studies.

The closer look analysis of Fig.\,2(b,c) and Fig.\,S5 of the Supplementary material~\cite{SUPL} shows that around ``strongly'' bonded regions of graphene on Rh(111) ($BRIDGE$, $HCP$, $FCC$) the full unit cell of graphene is imaged when only every second atom in the graphene unit cell is visible [larger hexagons in Fig.\,2(b,c)]. This is a well known effect for the STM or AFM imaging of the graphite- or graphene-based systems. This effect is also clearly visible in our STM images (see discussion above and Fig.\,1) analogous to the STM imaging of the graphene/Ni(111) system~\cite{Dedkov:2010a,Dzemiantsova:2011}. Increasing the frequency shift when one goes from image (b) to (c) leads to the increasing the imaging contrast and to the appearance of the atomically-resolved contrast above $ATOP$ positions, where now every carbon atom in the graphene unit cell can be resolved [smaller hexagon in Fig.\,2(c)]. This is a first observation of the fully resolved carbon ring of the strongly corrugated graphene layer on the metallic surface. The present effect can appear due to the local decoupling of graphene from the metallic substrate where the local hybridization between electron states of the graphene layer and the substrate is very small. A weak hybridization (if any) is also reflected in the carbon-atom-projected partial density of states (PDOS) and the difference of electron density at the graphene/Rh(111) interface (see also Figs.\,S6 and S7 of the Supplementary material~\cite{SUPL}). The corresponding PDOS for the carbon atoms around the $ATOP$ position is very close to the one for the free-standing graphene for which the imaging of every single carbon atom in the unit cell was obtained~\cite{Lauffer:2008}. 

\textit{In conclusion}, we present the studies of the graphene/Rh(111) system by means of DFT, STM, AFM, and force/tunneling current spectroscopy. The STM results show the perfect agreement with the theoretically calculated images and no principal difference in the contrast was found upon changing the sign of the bias voltage between the tip and the sample. On the contrary, the atomically-resolved AFM results demonstrate the dependence on the imaging conditions, namely on the frequency shift of the sensor. The obtained AFM and STM results on graphene/Rh(111) are understood on the basis of the DFT-D2 calculations and they demonstrate the high perspectives of AFM for the imaging of the chemical contrast in the heterogeneous systems.

E.\,N.\,V. acknowledges support from the DFG through the Collaborative Research Center (SFB) 765 and computer time at the North-German Supercomputing Alliance (HLRN). M.\,F. gratefully acknowledges the financial support by the European Science Foundation (ESF) under the EUROCORES Programme EuroGRAPHENE (Project ``SpinGraph'') and by the Research Center ``UltraQuantum'' (Excellence Initiative).



\clearpage

{\bf Figure captions:}
\newline
\newline
{\bf Fig.\,1.} (Color online) (a) Crystallographic structure and (b) the corresponding calculated STM image of graphene/Rh(111). (c) Large scale STM image of the graphene layer on Rh(111). Tunneling conditions: $U_T=+1$\,V, $I_T=1$\,nA. (d) 3D view of the $5\times5$\,nm$^2$ region from (c) showing the structure of the graphene layer on Rh(111) (see text for detailed discussion). Tunneling conditions: $U_T=-0.55$\,V, $I_T=10$\,nA.
\newline
\newline
{\bf Fig.\,2.} (Color online) (Upper panel) Frequency shift plots as functions of the relative distance with respect to the scanning position ($U_T=-840$\,mV, $I_T=0.48$\,nA) for \textit{hills} (A) and \textit{valleys} (B). The inset shows the respective tunneling current, $I(z)$. (Lower panel) The AFM images measured at the frequency shifts of the sensor marked by the corresponding dashed lines in the upper panel. The large and small hexagons mark the graphene unit cell and the carbon ring, respectively. The distorted rhombuses mark the graphene moir\'e cell on Rh(111).

\clearpage
\begin{figure}
\center
\includegraphics[scale=0.45]{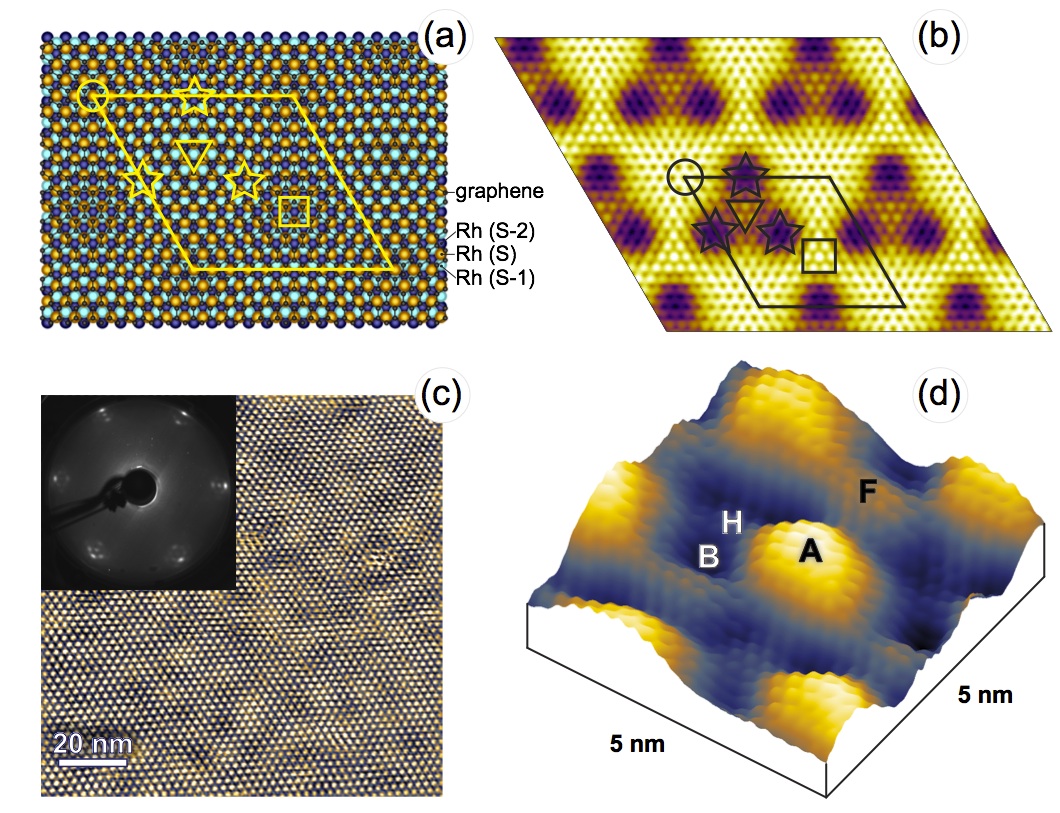}\\
\label{model-stm}
\vspace{1cm}
\large \textbf{Fig.\,1, E. N. Voloshina \textit{et al.}}
\end{figure}

\clearpage
\begin{figure}
\center
\includegraphics[scale=0.4]{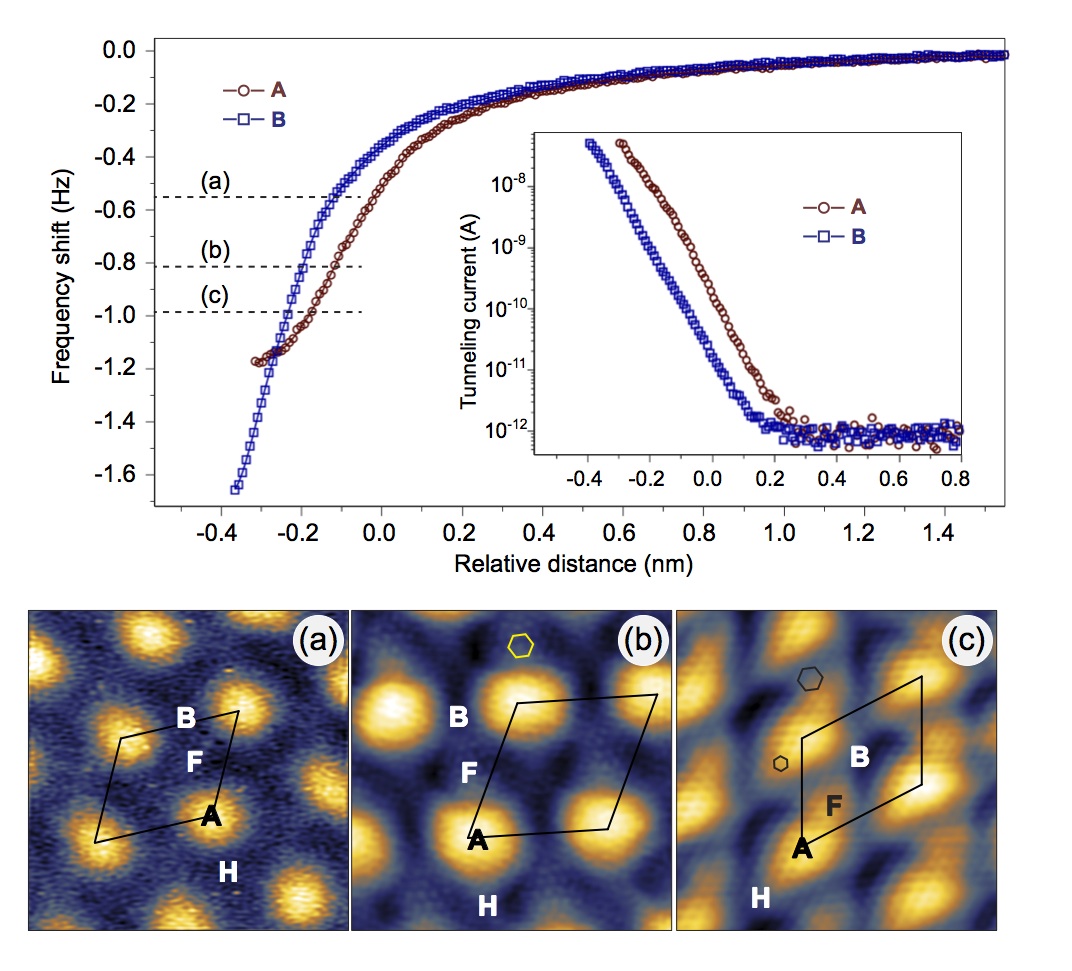}\\
\vspace{1cm}
\large \textbf{Fig.\,2, E. N. Voloshina \textit{et al.}}
\end{figure}

\clearpage

\noindent
Supplementary material for manuscript:\\
\textbf{Graphene on Rh(111): STM vs AFM studies}\\
\newline
E. N. Voloshina,$^1$ Yu. S. Dedkov,$^{2}$ S. Torbr\"ugge,$^{2}$ A. Thissen,$^{2}$ and M. Fonin$^3$\\
\newline
$^1$Physikalische und Theoretische Chemie, Freie Universit\"at Berlin, 14195 Berlin, Germany\\
$^2$SPECS Surface Nano Analysis GmbH, Voltastra\ss e 5, 13355 Berlin, Germany\\
$^3$Fachbereich Physik, Universit\"at Konstanz, 78457 Konstanz, Germany\\
\newline
\textbf{List of figures:}
\newline
\noindent\textbf{Fig.\,S1.} The height variation of the carbon atoms in the graphene layer on Rh(111). The corresponding high-symmetry positions are marked in the figure. The color bar on the right-hand side represents the hight scale in \AA.
\newline
\noindent\textbf{Fig.\,S2.} Large area atomically resolved experimental STM image (top view and 3D view) of graphene/Rh(111) demonstrating moir\'e structure. The symbols on the left-hand side image are the same as in Fig.\,1(a,b) of the manuscript.
\newline
\noindent\textbf{Fig.\,S3.} Calculated STM images of graphene/Rh(111) for the distance between the tip and the sample of $z=2$\,\AA. Integration was performed for the valence band stated in the energy region of $E-E_F=0 ...-1$\,eV (left panel) and $E-E_F=0 ...+1$\,eV (right panel).
\newline
\noindent\textbf{Fig.\,S4.} Corresponding force curves, $F(z)$, calculated on the basis of experimental data according to the formula presented in J. E. Sader and S. P. Jarvis, Appl. Phys. Lett. \textbf{84}, 1801 (2004).
\newline
\noindent\textbf{Fig.\,S5.} AFM image of graphene/Rh(111): the same as in Fig.\,2(c) for better view of the effects discussed in the manuscript. Every carbon atom is resolved in the graphene layer around the $ATOP$ position.
\newline
\noindent\textbf{Fig.\,S6.} Carbon atom-projected total density of states ($\sigma$ and $\pi$) in the valence band for the different high-symmetry positions of the graphene/Rh(111) system. The inset shows the corresponding density of states for the $p_z$ character only.
\newline
\noindent\textbf{Fig.\,S7.} Difference electron density, $\Delta n(r)=n_\mathrm{gr/Rh}(r)-n_\mathrm{Rh}(r)-n_\mathrm{gr}(r)$, plots in units of $e/\mathrm{\AA}^{3}$ calculated for graphene/Rh(111). Red (blue) colors indicate regions where the electron density increases (decreases). The images are made by means of VESTA visualization software [K. Momma and F. Izumi, ``VESTA 3 for three-dimensional visualization of crystal, volumetric and morphology data'', J. Appl. Crystallogr. \textbf{44}, 1272 (2011)].

\clearpage
\begin{figure}
\includegraphics[width=0.65\textwidth]{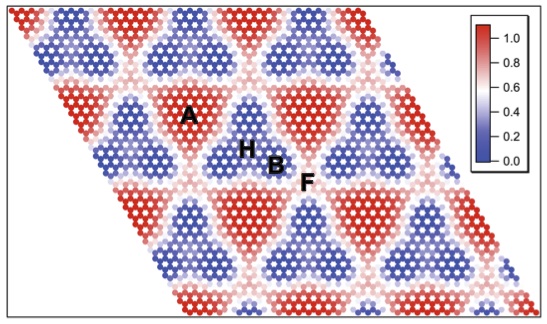}
\end{figure}
\noindent\textbf{Fig.\,S1.} The height variation of the carbon atoms in the graphene layer on Rh(111). The corresponding high-symmetry positions are marked in the figure. The color bar on the right-hand side represents the hight scale in \AA.

\clearpage
\begin{figure}
\includegraphics[width=0.75\textwidth]{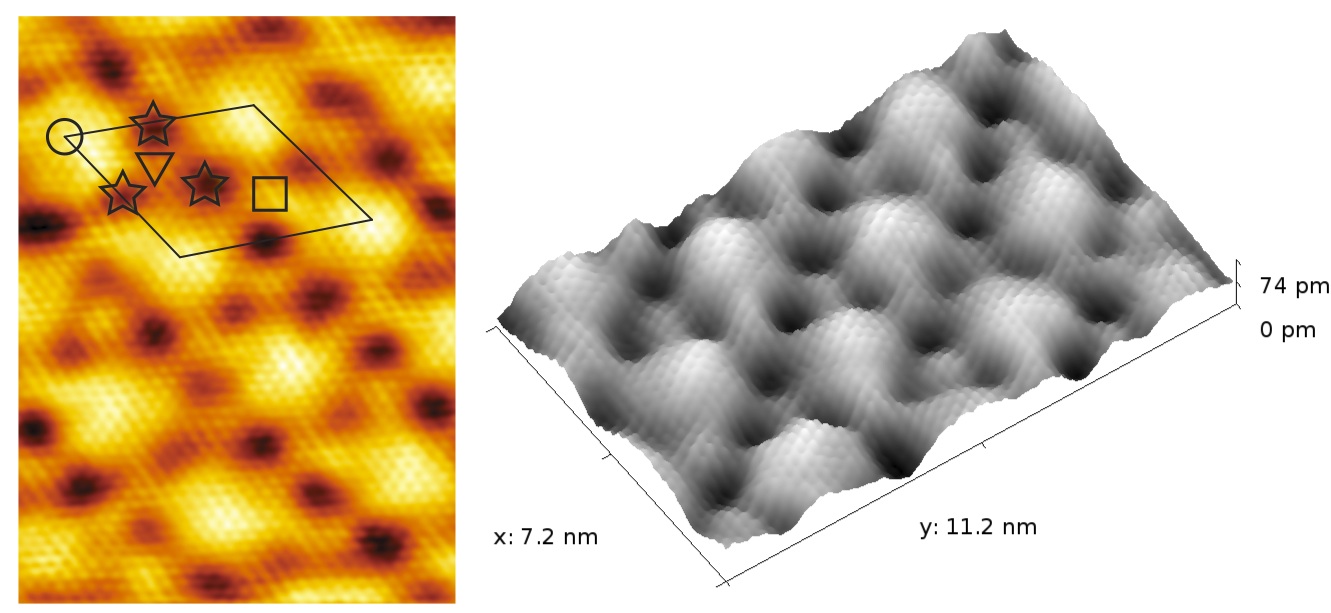}
\end{figure}
\noindent\textbf{Fig.\,S2.} Large area atomically resolved experimental STM image (top view and 3D view) of graphene/Rh(111) demonstrating moir\'e structure. The symbols on the left-hand side image are the same as in Fig.\,1(a,b) of the manuscript.

\clearpage
\begin{figure}
\includegraphics[width=0.8\textwidth]{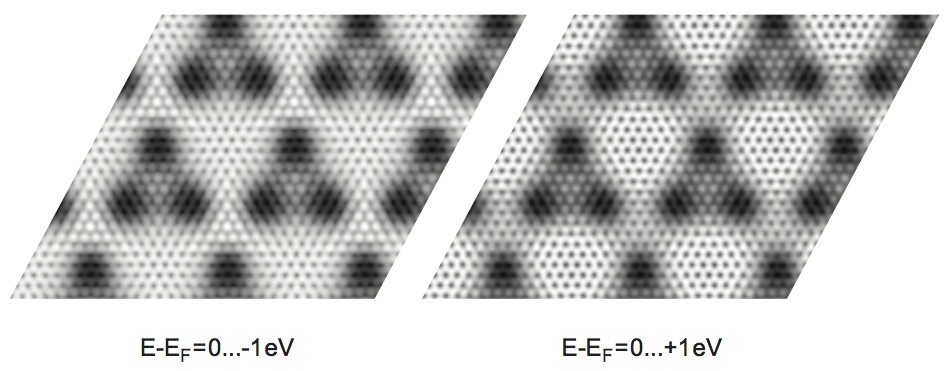}
\end{figure}
\noindent\textbf{Fig.\,S3.} Calculated STM images of graphene/Rh(111) for the distance between the tip and the sample of $z=2$\,\AA. Integration was performed for the valence band stated in the energy region of $E-E_F=0 ...-1$\,eV (left panel) and $E-E_F=0 ...+1$\,eV (right panel).

\clearpage
\begin{figure}
\includegraphics[width=0.75\textwidth]{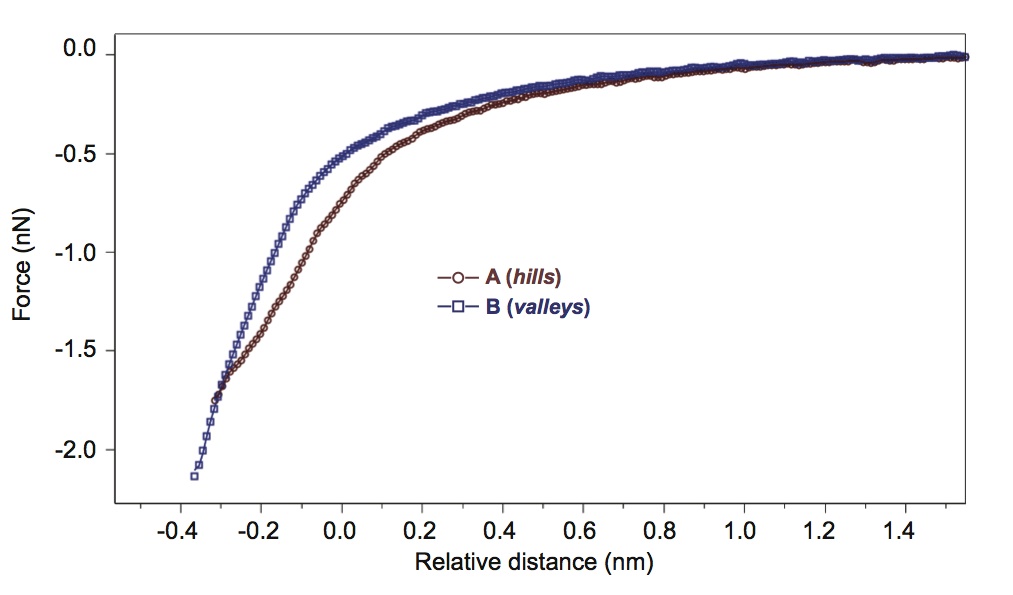}
\end{figure}
\noindent\textbf{Fig.\,S4.} Corresponding force curves, $F(z)$, calculated on the basis of experimental data according to the formula presented in J. E. Sader and S. P. Jarvis, Appl. Phys. Lett. \textbf{84}, 1801 (2004).

\clearpage
\begin{figure}
\includegraphics[width=1.0\textwidth]{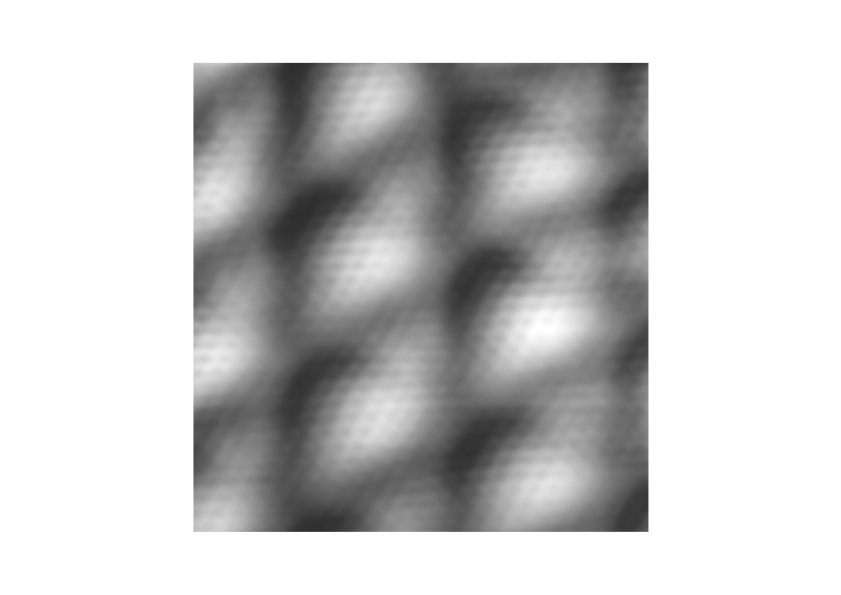}
\end{figure}
\noindent\textbf{Fig.\,S5.} AFM image of graphene/Rh(111): the same as in Fig.\,2(c) for better view of the effects discussed in the manuscript. Every carbon atom is resolved in the graphene layer around the $ATOP$ position.

\clearpage
\begin{figure}
\includegraphics[width=0.75\textwidth]{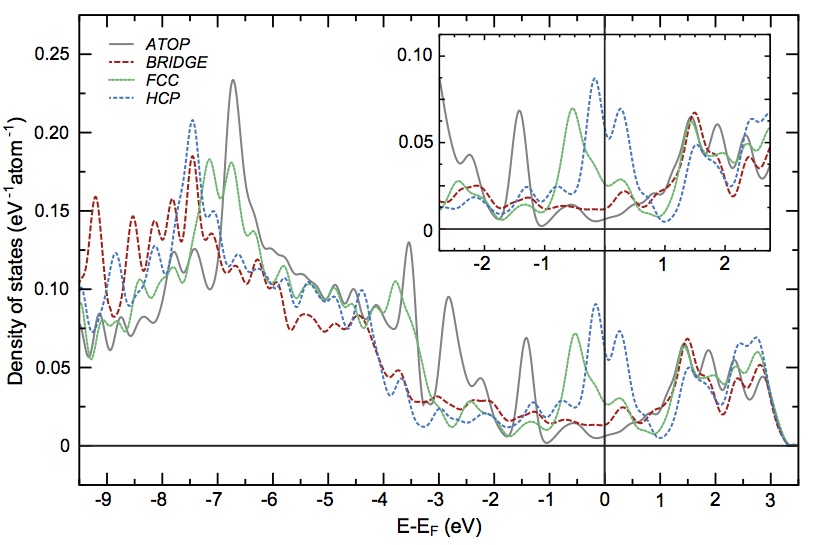}
\end{figure}
\noindent\textbf{Fig.\,S6.} Carbon atom-projected total density of states ($\sigma$ and $\pi$) in the valence band for the different high-symmetry positions of the graphene/Rh(111) system. The inset shows the corresponding density of states for the $p_z$ character only.

\clearpage
\begin{figure}
\includegraphics[width=0.75\textwidth]{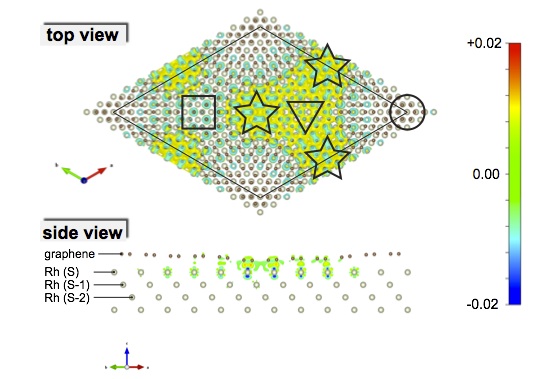}
\end{figure}
\noindent\textbf{Fig.\,S7.} Difference electron density, $\Delta n(r)=n_\mathrm{gr/Rh}(r)-n_\mathrm{Rh}(r)-n_\mathrm{gr}(r)$, plots in units of $e/\mathrm{\AA}^{3}$ calculated for graphene/Rh(111). Red (blue) colors indicate regions where the electron density increases (decreases). The images are made by means of VESTA visualization software [K. Momma and F. Izumi, ``VESTA 3 for three-dimensional visualization of crystal, volumetric and morphology data'', J. Appl. Crystallogr. \textbf{44}, 1272 (2011)].


\begin{thebibliography}{31}
\expandafter\ifx\csname natexlab\endcsname\relax\def\natexlab#1{#1}\fi
\expandafter\ifx\csname bibnamefont\endcsname\relax
  \def\bibnamefont#1{#1}\fi
\expandafter\ifx\csname bibfnamefont\endcsname\relax
  \def\bibfnamefont#1{#1}\fi
\expandafter\ifx\csname citenamefont\endcsname\relax
  \def\citenamefont#1{#1}\fi
\expandafter\ifx\csname url\endcsname\relax
  \def\url#1{\texttt{#1}}\fi
\expandafter\ifx\csname urlprefix\endcsname\relax\def\urlprefix{URL }\fi
\providecommand{\bibinfo}[2]{#2}
\providecommand{\eprint}[2][]{\url{#2}}

\bibitem[{\citenamefont{Geim}(2009)}]{Geim:2009}
\bibinfo{author}{\bibfnamefont{A.}~\bibnamefont{Geim}},
  \bibinfo{journal}{Science} \textbf{\bibinfo{volume}{324}},
  \bibinfo{pages}{1530} (\bibinfo{year}{2009}).

\bibitem[{\citenamefont{Geim}(2011)}]{Geim:2011}
\bibinfo{author}{\bibfnamefont{A.}~\bibnamefont{Geim}},
  \bibinfo{journal}{Rev. Mod. Phys.} \textbf{\bibinfo{volume}{83}},
  \bibinfo{pages}{851} (\bibinfo{year}{2011}).

\bibitem[{\citenamefont{Novoselov}(2011)}]{Novoselov:2011}
\bibinfo{author}{\bibfnamefont{K.}~\bibnamefont{Novoselov}},
  \bibinfo{journal}{Rev. Mod. Phys.} \textbf{\bibinfo{volume}{83}},
  \bibinfo{pages}{837} (\bibinfo{year}{2011}).

\bibitem[{\citenamefont{Schedin et~al.}(2007)\citenamefont{Schedin, Geim,
  Morozov, Hill, Blake, Katsnelson, and Novoselov}}]{Schedin:2007}
\bibinfo{author}{\bibfnamefont{F.}~\bibnamefont{Schedin}},
  \bibinfo{author}{\bibfnamefont{A.~K.} \bibnamefont{Geim}},
  \bibinfo{author}{\bibfnamefont{S.~V.} \bibnamefont{Morozov}},
  \bibinfo{author}{\bibfnamefont{E.~W.} \bibnamefont{Hill}},
  \bibinfo{author}{\bibfnamefont{P.}~\bibnamefont{Blake}},
  \bibinfo{author}{\bibfnamefont{M.~I.} \bibnamefont{Katsnelson}},
  \bibnamefont{and} \bibinfo{author}{\bibfnamefont{K.~S.}
  \bibnamefont{Novoselov}}, \bibinfo{journal}{Nature Mater.}
  \textbf{\bibinfo{volume}{6}}, \bibinfo{pages}{652} (\bibinfo{year}{2007}).

\bibitem[{\citenamefont{Lin et~al.}(2010)\citenamefont{Lin, Dimitrakopoulos,
  Jenkins, Farmer, Chiu, Grill, and Avouris}}]{Lin:2010}
\bibinfo{author}{\bibfnamefont{Y.~M.} \bibnamefont{Lin}},
  \bibinfo{author}{\bibfnamefont{C.}~\bibnamefont{Dimitrakopoulos}},
  \bibinfo{author}{\bibfnamefont{K.~A.} \bibnamefont{Jenkins}},
  \bibinfo{author}{\bibfnamefont{D.~B.} \bibnamefont{Farmer}},
  \bibinfo{author}{\bibfnamefont{H.~Y.} \bibnamefont{Chiu}},
  \bibinfo{author}{\bibfnamefont{A.}~\bibnamefont{Grill}}, \bibnamefont{and}
  \bibinfo{author}{\bibfnamefont{P.}~\bibnamefont{Avouris}},
  \bibinfo{journal}{Science} \textbf{\bibinfo{volume}{327}},
  \bibinfo{pages}{662} (\bibinfo{year}{2010}).

\bibitem[{\citenamefont{Lin et~al.}(2011)\citenamefont{Lin, Valdes-Garcia, Han,
  Farmer, Meric, Sun, Wu, Dimitrakopoulos, Grill, Avouris et~al.}}]{Lin:2011}
\bibinfo{author}{\bibfnamefont{Y.-M.} \bibnamefont{Lin}},
  \bibinfo{author}{\bibfnamefont{A.}~\bibnamefont{Valdes-Garcia}},
  \bibinfo{author}{\bibfnamefont{S.-J.} \bibnamefont{Han}},
  \bibinfo{author}{\bibfnamefont{D.~B.} \bibnamefont{Farmer}},
  \bibinfo{author}{\bibfnamefont{I.}~\bibnamefont{Meric}},
  \bibinfo{author}{\bibfnamefont{Y.}~\bibnamefont{Sun}},
  \bibinfo{author}{\bibfnamefont{Y.}~\bibnamefont{Wu}},
  \bibinfo{author}{\bibfnamefont{C.}~\bibnamefont{Dimitrakopoulos}},
  \bibinfo{author}{\bibfnamefont{A.}~\bibnamefont{Grill}},
  \bibinfo{author}{\bibfnamefont{P.}~\bibnamefont{Avouris}}
  \bibnamefont{et~al.}, \bibinfo{journal}{Science}
  \textbf{\bibinfo{volume}{332}}, \bibinfo{pages}{1294} (\bibinfo{year}{2011}).

\bibitem[{\citenamefont{Bae et~al.}(2010)\citenamefont{Bae, Kim, Lee, Xu, Park,
  Zheng, Balakrishnan, Lei, Kim, Song et~al.}}]{Bae:2010}
\bibinfo{author}{\bibfnamefont{S.}~\bibnamefont{Bae}},
  \bibinfo{author}{\bibfnamefont{H.}~\bibnamefont{Kim}},
  \bibinfo{author}{\bibfnamefont{Y.}~\bibnamefont{Lee}},
  \bibinfo{author}{\bibfnamefont{X.}~\bibnamefont{Xu}},
  \bibinfo{author}{\bibfnamefont{J.-S.} \bibnamefont{Park}},
  \bibinfo{author}{\bibfnamefont{Y.}~\bibnamefont{Zheng}},
  \bibinfo{author}{\bibfnamefont{J.}~\bibnamefont{Balakrishnan}},
  \bibinfo{author}{\bibfnamefont{T.}~\bibnamefont{Lei}},
  \bibinfo{author}{\bibfnamefont{H.~R.} \bibnamefont{Kim}},
  \bibinfo{author}{\bibfnamefont{Y.~I.} \bibnamefont{Song}}
  \bibnamefont{et~al.}, \bibinfo{journal}{Nature Nanotech.}
  \textbf{\bibinfo{volume}{5}}, \bibinfo{pages}{574} (\bibinfo{year}{2010}).

\bibitem[{\citenamefont{Wintterlin and Bocquet}(2009)}]{Wintterlin:2009}
\bibinfo{author}{\bibfnamefont{J.}~\bibnamefont{Wintterlin}} \bibnamefont{and}
  \bibinfo{author}{\bibfnamefont{M.~L.} \bibnamefont{Bocquet}},
  \bibinfo{journal}{Surf. Sci.} \textbf{\bibinfo{volume}{603}},
  \bibinfo{pages}{1841} (\bibinfo{year}{2009}).

\bibitem[{\citenamefont{Batzill}(2012)}]{Batzill:2012}
\bibinfo{author}{\bibfnamefont{M.}~\bibnamefont{Batzill}},
  \bibinfo{journal}{Surf. Sci. Rep.} \textbf{\bibinfo{volume}{67}},
  \bibinfo{pages}{83} (\bibinfo{year}{2012}).

\bibitem[{\citenamefont{Dedkov et~al.}()\citenamefont{Dedkov, Horn,
  Preobrajenskij, and Fonin}}]{Dedkov:2012book}
\bibinfo{author}{\bibfnamefont{Yu.~S.} \bibnamefont{Dedkov}},
  \bibinfo{author}{\bibfnamefont{K.}~\bibnamefont{Horn}},
  \bibinfo{author}{\bibfnamefont{A.}~\bibnamefont{Preobrajenskij}},
  \bibnamefont{and} \bibinfo{author}{\bibfnamefont{M.}~\bibnamefont{Fonin}}, in
  \emph{\bibinfo{booktitle}{Graphene Nanoelectronics}}, ed.
  \bibinfo{editor}{\bibfnamefont{H.}~\bibnamefont{Raza}}
  (\bibinfo{publisher}{Springer}, \bibinfo{address}{Berlin}, 2012).

\bibitem[{\citenamefont{Dedkov et~al.}(2008{\natexlab{a}})\citenamefont{Dedkov,
  Fonin, and Laubschat}}]{Dedkov:2008d}
\bibinfo{author}{\bibfnamefont{Yu.~S.} \bibnamefont{Dedkov}},
  \bibinfo{author}{\bibfnamefont{M.}~\bibnamefont{Fonin}}, \bibnamefont{and}
  \bibinfo{author}{\bibfnamefont{C.}~\bibnamefont{Laubschat}},
  \bibinfo{journal}{Appl. Phys. Lett.} \textbf{\bibinfo{volume}{92}},
  \bibinfo{pages}{052506} (\bibinfo{year}{2008}{\natexlab{a}}).

\bibitem[{\citenamefont{Dedkov et~al.}(2008{\natexlab{b}})\citenamefont{Dedkov,
  Fonin, Ruediger, and Laubschat}}]{Dedkov:2008e}
\bibinfo{author}{\bibfnamefont{Yu.~S.} \bibnamefont{Dedkov}},
  \bibinfo{author}{\bibfnamefont{M.}~\bibnamefont{Fonin}},
  \bibinfo{author}{\bibfnamefont{U.}~\bibnamefont{R\"udiger}}, \bibnamefont{and}
  \bibinfo{author}{\bibfnamefont{C.}~\bibnamefont{Laubschat}},
  \bibinfo{journal}{Appl. Phys. Lett.} \textbf{\bibinfo{volume}{93}},
  \bibinfo{pages}{022509} (\bibinfo{year}{2008}{\natexlab{b}}).

\bibitem[{\citenamefont{Sutter et~al.}(2010)\citenamefont{Sutter, Albrecht,
  Camino, and Sutter}}]{Sutter:2010bx}
\bibinfo{author}{\bibfnamefont{E.}~\bibnamefont{Sutter}},
  \bibinfo{author}{\bibfnamefont{P.}~\bibnamefont{Albrecht}},
  \bibinfo{author}{\bibfnamefont{F.~E.} \bibnamefont{Camino}},
  \bibnamefont{and} \bibinfo{author}{\bibfnamefont{P.}~\bibnamefont{Sutter}},
  \bibinfo{journal}{Carbon} \textbf{\bibinfo{volume}{48}},
  \bibinfo{pages}{4414} (\bibinfo{year}{2010}{\natexlab{b}}).

\bibitem[{\citenamefont{Chen et~al.}(2011)\citenamefont{Chen, Brown, Levendorf,
  Cai, Ju, Edgeworth, Li, Magnuson, Velamakanni, Piner et~al.}}]{Chen:2011a}
\bibinfo{author}{\bibfnamefont{S.}~\bibnamefont{Chen}},
  \bibinfo{author}{\bibfnamefont{L.}~\bibnamefont{Brown}},
  \bibinfo{author}{\bibfnamefont{M.}~\bibnamefont{Levendorf}},
  \bibinfo{author}{\bibfnamefont{W.}~\bibnamefont{Cai}},
  \bibinfo{author}{\bibfnamefont{S.-Y.} \bibnamefont{Ju}},
  \bibinfo{author}{\bibfnamefont{J.}~\bibnamefont{Edgeworth}},
  \bibinfo{author}{\bibfnamefont{X.}~\bibnamefont{Li}},
  \bibinfo{author}{\bibfnamefont{C.~W.} \bibnamefont{Magnuson}},
  \bibinfo{author}{\bibfnamefont{A.}~\bibnamefont{Velamakanni}},
  \bibinfo{author}{\bibfnamefont{R.~D.} \bibnamefont{Piner}}
  \bibnamefont{et~al.}, \bibinfo{journal}{ACS Nano}
  \textbf{\bibinfo{volume}{5}}, \bibinfo{pages}{1321} (\bibinfo{year}{2011}).

\bibitem[{\citenamefont{Karpan et~al.}(2007)\citenamefont{Karpan, Giovannetti,
  Khomyakov, Talanana, Starikov, Zwierzycki, Brink, Brocks, and
  Kelly}}]{Karpan:2007}
\bibinfo{author}{\bibfnamefont{V.~M.} \bibnamefont{Karpan}},
  \bibinfo{author}{\bibfnamefont{G.}~\bibnamefont{Giovannetti}},
  \bibinfo{author}{\bibfnamefont{P.~A.} \bibnamefont{Khomyakov}},
  \bibinfo{author}{\bibfnamefont{M.}~\bibnamefont{Talanana}},
  \bibinfo{author}{\bibfnamefont{A.~A.} \bibnamefont{Starikov}},
  \bibinfo{author}{\bibfnamefont{M.}~\bibnamefont{Zwierzycki}},
  \bibinfo{author}{\bibfnamefont{J.}~\bibnamefont{van der Brink}},
  \bibinfo{author}{\bibfnamefont{G.}~\bibnamefont{Brocks}}, \bibnamefont{and}
  \bibinfo{author}{\bibfnamefont{P.~J.} \bibnamefont{Kelly}},
  \bibinfo{journal}{Phys. Rev. Lett.} \textbf{\bibinfo{volume}{99}},
  \bibinfo{pages}{176602} (\bibinfo{year}{2007}).

\bibitem[{\citenamefont{Karpan et~al.}(2008)\citenamefont{Karpan, Khomyakov,
  Starikov, Giovannetti, Zwierzycki, Talanana, Brocks, Brink, and
  Kelly}}]{Karpan:2008}
\bibinfo{author}{\bibfnamefont{V.~M.} \bibnamefont{Karpan}},
  \bibinfo{author}{\bibfnamefont{P.~A.} \bibnamefont{Khomyakov}},
  \bibinfo{author}{\bibfnamefont{A.~A.} \bibnamefont{Starikov}},
  \bibinfo{author}{\bibfnamefont{G.}~\bibnamefont{Giovannetti}},
  \bibinfo{author}{\bibfnamefont{M.}~\bibnamefont{Zwierzycki}},
  \bibinfo{author}{\bibfnamefont{M.}~\bibnamefont{Talanana}},
  \bibinfo{author}{\bibfnamefont{G.}~\bibnamefont{Brocks}},
  \bibinfo{author}{\bibfnamefont{J.}~\bibnamefont{van der Brink}},
  \bibnamefont{and} \bibinfo{author}{\bibfnamefont{P.~J.} \bibnamefont{Kelly}},
  \bibinfo{journal}{Phys. Rev. B} \textbf{\bibinfo{volume}{78}},
  \bibinfo{pages}{195419} (\bibinfo{year}{2008}).

\bibitem[{\citenamefont{Dedkov and Fonin}(2010)}]{Dedkov:2010a}
\bibinfo{author}{\bibfnamefont{Yu.~S.} \bibnamefont{Dedkov}} \bibnamefont{and}
  \bibinfo{author}{\bibfnamefont{M.}~\bibnamefont{Fonin}},
  \bibinfo{journal}{New J. Phys.} \textbf{\bibinfo{volume}{12}},
  \bibinfo{pages}{125004} (\bibinfo{year}{2010}).

\bibitem[{\citenamefont{N'Diaye et~al.}(2009)\citenamefont{N'Diaye, Gerber,
  Busse, Myslivecek, Coraux, and Michely}}]{NDiaye:2009a}
\bibinfo{author}{\bibfnamefont{A.~T.} \bibnamefont{N'Diaye}},
  \bibinfo{author}{\bibfnamefont{T.}~\bibnamefont{Gerber}},
  \bibinfo{author}{\bibfnamefont{C.}~\bibnamefont{Busse}},
  \bibinfo{author}{\bibfnamefont{J.}~\bibnamefont{Myslivecek}},
  \bibinfo{author}{\bibfnamefont{J.}~\bibnamefont{Coraux}}, \bibnamefont{and}
  \bibinfo{author}{\bibfnamefont{T.}~\bibnamefont{Michely}},
  \bibinfo{journal}{New J. Phys.} \textbf{\bibinfo{volume}{11}},
  \bibinfo{pages}{103045} (\bibinfo{year}{2009}).

\bibitem[{\citenamefont{Sicot et~al.}(2010)\citenamefont{Sicot, Bouvron,
  Zander, Ruediger, Dedkov, and Fonin}}]{Sicot:2010}
\bibinfo{author}{\bibfnamefont{M.}~\bibnamefont{Sicot}},
  \bibinfo{author}{\bibfnamefont{S.}~\bibnamefont{Bouvron}},
  \bibinfo{author}{\bibfnamefont{O.}~\bibnamefont{Zander}},
  \bibinfo{author}{\bibfnamefont{U.}~\bibnamefont{R\"udiger}},
  \bibinfo{author}{\bibfnamefont{Yu.~S.} \bibnamefont{Dedkov}},
  \bibnamefont{and} \bibinfo{author}{\bibfnamefont{M.}~\bibnamefont{Fonin}},
  \bibinfo{journal}{Appl. Phys. Lett.} \textbf{\bibinfo{volume}{96}},
  \bibinfo{pages}{093115} (\bibinfo{year}{2010}).

\bibitem{Sugimoto:2007} Y. Sugimoto, P. Pou, M. Abe, P. Jelinek, R. P\'erez, S. Morita, and \'O. Custance, Nature \textbf{446}, 64 (2007).

\bibitem[{\citenamefont{BLOCHL}(1994)}]{Blochl:1994}
\bibinfo{author}{\bibfnamefont{P.}~\bibnamefont{Blochl}},
  \bibinfo{journal}{Phys. Rev. B} \textbf{\bibinfo{volume}{50}},
  \bibinfo{pages}{17953} (\bibinfo{year}{1994}).

\bibitem[{\citenamefont{Perdew et~al.}(1996)\citenamefont{Perdew, Burke, and
  Ernzerhof}}]{Perdew:1996}
\bibinfo{author}{\bibfnamefont{J.}~\bibnamefont{Perdew}},
  \bibinfo{author}{\bibfnamefont{K.}~\bibnamefont{Burke}}, \bibnamefont{and}
  \bibinfo{author}{\bibfnamefont{M.}~\bibnamefont{Ernzerhof}},
  \bibinfo{journal}{Phys. Rev. Lett.} \textbf{\bibinfo{volume}{77}},
  \bibinfo{pages}{3865} (\bibinfo{year}{1996}).

\bibitem[{\citenamefont{Kresse and Hafner}(1994)}]{Kresse:1994}
\bibinfo{author}{\bibfnamefont{G.}~\bibnamefont{Kresse}} \bibnamefont{and}
  \bibinfo{author}{\bibfnamefont{J.}~\bibnamefont{Hafner}}, \bibinfo{journal}{J.
  Phys.: Condens. Matter} \textbf{\bibinfo{volume}{6}}, \bibinfo{pages}{8245}
  (\bibinfo{year}{1994}).

\bibitem[{\citenamefont{Grimme}(2006)}]{Grimme:2006}
\bibinfo{author}{\bibfnamefont{S.}~\bibnamefont{Grimme}}, \bibinfo{journal}{J.
  Comput. Chem.} \textbf{\bibinfo{volume}{27}}, \bibinfo{pages}{1787}
  (\bibinfo{year}{2006}).

\bibitem[{\citenamefont{NEUGEBAUER and SCHEFFLER}(1992)}]{Neugebauer:1992}
\bibinfo{author}{\bibfnamefont{J.}~\bibnamefont{Neugebauer}} \bibnamefont{and}
  \bibinfo{author}{\bibfnamefont{M.}~\bibnamefont{Scheffler}},
  \bibinfo{journal}{Phys. Rev. B} \textbf{\bibinfo{volume}{46}},
  \bibinfo{pages}{16067} (\bibinfo{year}{1992}).

\bibitem[{\citenamefont{TERSOFF and HAMANN}(1985)}]{Tersoff:1985}
\bibinfo{author}{\bibfnamefont{J.}~\bibnamefont{Tersoff}} \bibnamefont{and}
  \bibinfo{author}{\bibfnamefont{D.}~\bibnamefont{Hamann}},
  \bibinfo{journal}{Phys. Rev. B} \textbf{\bibinfo{volume}{31}},
  \bibinfo{pages}{805} (\bibinfo{year}{1985}).

\bibitem[{\citenamefont{Vanpoucke and Brocks}(2008)}]{VanPoucke:2008}
\bibinfo{author}{\bibfnamefont{D.~E.~P.} \bibnamefont{Vanpoucke}}
  \bibnamefont{and} \bibinfo{author}{\bibfnamefont{G.}~\bibnamefont{Brocks}},
  \bibinfo{journal}{Phys. Rev. B} \textbf{\bibinfo{volume}{77}},
  \bibinfo{pages}{241308} (\bibinfo{year}{2008}).

\bibitem[{\citenamefont{Voloshina et~al.}(2011)\citenamefont{Voloshina,
  Generalov, Weser, B{\"o}ttcher, Horn, and Dedkov}}]{Voloshina:2011NJP}
\bibinfo{author}{\bibfnamefont{E.~N.} \bibnamefont{Voloshina}},
  \bibinfo{author}{\bibfnamefont{A.}~\bibnamefont{Generalov}},
  \bibinfo{author}{\bibfnamefont{M.}~\bibnamefont{Weser}},
  \bibinfo{author}{\bibfnamefont{S.}~\bibnamefont{B{\"o}ttcher}},
  \bibinfo{author}{\bibfnamefont{K.}~\bibnamefont{Horn}}, \bibnamefont{and}
  \bibinfo{author}{\bibfnamefont{Yu.~S.} \bibnamefont{Dedkov}},
  \bibinfo{journal}{New J. Phys.} \textbf{\bibinfo{volume}{13}},
  \bibinfo{pages}{113028} (\bibinfo{year}{2011}).

\bibitem[{\citenamefont{Sicot et~al.}(2012)\citenamefont{Sicot, Leicht, Zusan,
  Bouvron, Zander, Weser, Dedkov, Horn, and Fonin}}]{Sicot:2012}
\bibinfo{author}{\bibfnamefont{M.}~\bibnamefont{Sicot}},
  \bibinfo{author}{\bibfnamefont{P.}~\bibnamefont{Leicht}},
  \bibinfo{author}{\bibfnamefont{A.}~\bibnamefont{Zusan}},
  \bibinfo{author}{\bibfnamefont{S.}~\bibnamefont{Bouvron}},
  \bibinfo{author}{\bibfnamefont{O.}~\bibnamefont{Zander}},
  \bibinfo{author}{\bibfnamefont{M.}~\bibnamefont{Weser}},
  \bibinfo{author}{\bibfnamefont{Yu.~S.} \bibnamefont{Dedkov}},
  \bibinfo{author}{\bibfnamefont{K.}~\bibnamefont{Horn}}, \bibnamefont{and}
  \bibinfo{author}{\bibfnamefont{M.}~\bibnamefont{Fonin}},
  \bibinfo{journal}{ACS Nano} \textbf{\bibinfo{volume}{6}},
  \bibinfo{pages}{151} (\bibinfo{year}{2012}).

\bibitem[{SPE()}]{SPECS}
{\bibinfo{title}{{http://www.specs.com}}}.

\bibitem{Torbrugge:2010}
S. Torbr\"ugge, O. Schaff, and J. Rychen, J. Vac. Sci. Technol. B \textbf{28}, C4E12 (2010).

\bibitem{SUPL}
{\bibinfo{title}{{See supplementary material at [URL will be inserted by AIP] for additional experimental and theoretical data as well as for details of analysis}}}.

\bibitem[{\citenamefont{Wang et~al.}(2010)\citenamefont{Wang, Caffio, Bromley,
  Fr{\"u}chtl, and Schaub}}]{Wang:2010ky}
\bibinfo{author}{\bibfnamefont{B.}~\bibnamefont{Wang}},
  \bibinfo{author}{\bibfnamefont{M.}~\bibnamefont{Caffio}},
  \bibinfo{author}{\bibfnamefont{C.}~\bibnamefont{Bromley}},
  \bibinfo{author}{\bibfnamefont{H.}~\bibnamefont{Fr{\"u}chtl}},
  \bibnamefont{and} \bibinfo{author}{\bibfnamefont{R.}~\bibnamefont{Schaub}},
  \bibinfo{journal}{ACS Nano} \textbf{\bibinfo{volume}{4}},
  \bibinfo{pages}{5773} (\bibinfo{year}{2010}).

\bibitem[{\citenamefont{Wang and Bocquet}(2011)}]{Wang:2011hh}
\bibinfo{author}{\bibfnamefont{B.}~\bibnamefont{Wang}} \bibnamefont{and}
  \bibinfo{author}{\bibfnamefont{M.-L.} \bibnamefont{Bocquet}},
  \bibinfo{journal}{J. Phys. Chem. Lett.} \textbf{\bibinfo{volume}{2}},
  \bibinfo{pages}{2341} (\bibinfo{year}{2011}).

\bibitem{Dzemiantsova:2011} L. V. Dzemiantsova, M. Karolak, F. Lofink, A. Kubetzka, B. Sachs, K. von Bergmann, S. Hankemeier, T. O. Wehling, R. Fr\"omter, H. P. Oepen, A. I. Lichtenstein, and R. Wiesendanger, Phys. Rev. B \textbf{84}, 205431 (2011).

\bibitem{Lauffer:2008} P. Lauffer, K. V. Emtsev, R. Graupner, Th. Seyller, L. Ley, S. A. Reshanov, and H. B. Weber, Phys. Rev. B \textbf{77}, 155426 (2008).

\end{thebibliography}
\end{document}